\documentclass[12pt]{iopart}

\usepackage{iopams}
\usepackage{graphicx}
\usepackage[breaklinks=true,colorlinks=true,linkcolor=blue,urlcolor=blue,citecolor=blue]{hyperref}
\begin{document}

\title[Heralded generation of single photons entangled in multiple temporal modes]{Heralded generation of single photons entangled in multiple temporal modes with controllable waveforms}

\author{A. Gogyan,$^{1*}$ N. Sisakyan, $^1$ R.  Akhmedzhanov $^2$ and Yu. Malakyan$^{1,3}$}

\address{$^{1}$ Institute for Physical Research, Armenian
National Academy of Sciences, Ashtarak-2, 0203, Armenia}
\address{$^{2}$ Institute of Applied Physics, Russian Academy
of Sciences,ul. Ul'yanova 46, Nizhni Novgorod, 603950 Russia}
\address{$^{3}$ Centre of Strong Field Physics, Yerevan State
University, 1 A. Manukian St., Yerevan 0025, Armenia}
\ead{$^{*}$agogyan@gmail.com}

\begin{abstract}
Time-bin entangled single-photons are highly demanded for long
distance quantum communication. We propose a heralded source of
tunable narrowband single photons entangled in well-separated
multiple temporal modes (time bins) with controllable amplitudes.
The detection of a single Stokes photon generated in a cold atomic
ensemble via Raman scattering of a weak write pulse heralds the
preparation of one spin excitation stored within the atomic
medium. A train of read laser pulses deterministically converts
the atomic excitation into a single anti-Stokes photon delocalized
in multi-time-bins. The waveforms of bins are well controlled by
the read pulse parameters. A scheme to measure the phase coherence
across all time bins is suggested.
\end{abstract}

\maketitle

\section{Introduction}
The research on generation of photon states entangled in
multi-temporal modes is rapidly growing recently. This interest is
due to two main reasons. First, this type of entanglement can be
transferred over significantly large distances with very little
decoherence \cite{1,2}, which allows much more robust quantum
communication systems in contrast to types of entanglement based
on polarization-encoded qubits. Second, employment of time-bin
entangled photons allows to perform linear optical quantum
computing (LOQC) in a single spatial mode \cite{3}, providing
scalable implementation of many-qubit protocols without creating
unwieldy networks which inevitably arise in all schemes of LOQC
\cite{4,5} due to many spatial modes.

Generation of optical pulses in distinct temporal modes has been
studied experimentally in a variety of settings. The
interferometric preparation of time-bin qubits of broadband
photons generated in non-linear crystals from a spontaneous
parametric down conversion (SPDC) has been reported in Refs.
\cite{6,7,8,9}. The conversion of weak signal pulses into
multi-temporal modes in room-temperature vapors has been shown
under electromagnetic induced transparency conditions \cite{10}
and far-off resonant Raman schemes \cite{11}. The generation of
time-bin entangled photon pairs in the $1.5\mu$m band via
spontaneous four-wave mixing in a cooled fiber is demonstrated in
\cite{12}. There exist different models to realize such photonic
states based, for example, on the parametric interaction between
two single-photon pulses in a coherent atomic ensemble \cite{13},
stimulated Raman adiabatic passage in a single quantum dot
\cite{14} or in an atom-cavity system \cite{15}. However, each of
these methods has certain disadvantages impeding their use as
efficient sources of time-bin entangled photons. The good source
should provide: i) a pure single-photon (SP) state without mixture
from both the multi-photon and zero-photon states; ii) well
separated quantum temporal modes such that one can perform
spacelike separated local measurements on the modes; iii)
controllable amplitudes of the temporal modes, and iv) phase
coherence across all time bins. The schemes proposed so far do not
fulfill all these requirements. In particular, for SPDC in
non-linear crystals, the generation of pure SP states and
controllability of the waveforms of temporal modes are the main
challenges. Besides, the photon linewidth is too broad to address
atomic transitions effectively. For atomic ensembles, the
repetition rate and a nonzero probability of generating more than
one photon are major obstacles. Also, in some models, the number
of SP temporal modes and delay between the bins are not easily
controlled.

In this paper we propose a scheme, which is a promising candidate
for high-quality sources of time-bin entangled SP states featuring
the properties listed above. The proposed scheme is a heralded
generation system, where at first, similar to DLCZ protocol
\cite{16}, a Stokes photon is produced via Raman scattering of a
weak write pulse in an ensemble of cold lambda-atoms confined
inside a hollow core of a single-mode photonic-crystal fiber
(HC-PCF) (Figure 1). The successful detection of the Stokes photon
by two single-photon detectors D1 and D2 heralds the creation of
one collective excitation in the atomic ensemble. Then, the atomic
excitation is converted into one anti-Stokes photon by applying a
train of phase-locked read laser pulses, the number and
intensities of which are adjusted such that an individual read
pulse cannot retrieve the anti-Stokes photon completely, but the
total conversion is highly efficient with the probability one.
This is achieved due to the fiber enhanced atom-photon interaction
and multiatom collective interference effects \cite{16}. As a
result, the anti-Stokes photon is emitted in a well-defined
spatial mode being coherently localized in many temporal modes.
Note that the control of anti-Stokes photon waveform by varying
the intensity and frequency of read pulse has been demonstrated
experimentally in DLCZ scheme \cite{balic,mendes}.

The main limitation of our scheme is the low heralding efficiency
due to low probability for Stokes-photon emission needed to
exclude the multi-atom events in the collective spin excitation.
The additional imperfections may result from Stokes photon losses,
when a heralded signal is present, but no Stokes photon is
detected due to detector inefficiencies. Therefore, the
experimental verification of heralded creation of atomic
excitation is necessary in order to assess the single-excitation
regime for each ensemble. A convenient parameter is the
cross-correlation function between the Stokes and anti-Stokes
photons, the large values of which under conditions of weak Stokes
generation indicate the presence of a single excitation in the
medium. This protocol has been recently realized in cold atoms
confined in a magneto-optical trap \cite{17,18} by performing
sequential write trials and heralding measurements. The
cross-correlation function temporal structure have been shown in
Ref. \cite{polyakov}. After a single collective atomic spin
excitation is heralded our system operates as a deterministic
source of anti-Stokes single-photons entangled in multi-time-bins.

Our model is described more detailed in the next Section, where
the anti-Stokes photon flux is calculated for different sets of
read pulses clearly demonstrating a well-defined dependence of
time-bin waveforms on the profiles of read pulses. The
experimental test to verify the coherence between anti-Stokes
temporal modes is discussed in Section 3. The results are
concluded in Section 4.

\section{Model and basic equations}

The model discussed in this paper is based on our earlier works
\cite{19,20}. It describes a cold ensemble of $N$ four-level
atoms, which are trapped inside a HC-PCF of small diameter $D$ and
length $L$ (see Figure 1). We consider the atoms are initially
prepared in the state $|1\rangle$ by optical pumping and are
strongly confined in the transverse direction inside the fiber
core that prevents atom-wall collisions \cite{21}. The write
$\omega_W$ laser field with peak Rabi frequency $\Omega_W$
interacts with the atoms at the transition $1 \rightarrow 3$ and
generates a single Stokes photon at the transition $3 \rightarrow
2$. Since in the forward direction all the atoms are identically
and strongly coupled to the Stokes photon \cite{16}, a
symmetrically distributed spin excitation is stored in the medium.
We employ the far off-resonant Raman configuration (Figure 1) with
large detuning $\Delta_W=\omega _{31}-\omega _{W}$ that makes the
system immune to the spontaneous losses into field modes other
than the Stokes mode, as well as to dephasing effects induced by
other excited states. After a Stokes photon was successfully
detected by the D1:D2 block in Figure 1, a train of read
$\omega_R$ laser pulses is applied at the transition $2\rightarrow
4$ in a far off-resonant Raman configuration
$\Delta_R=\omega_{42}-\omega_{R}\neq 0$ with adjusted intensities
such that the atomic spin excitation is completely converted at
the transition $4 \rightarrow 1$ into a tunable anti-Stokes photon
delocalized in multi time-bins (the case of two read pulses is
shown in Figure 1) with amplitudes, which are easily controlled
being proportional to the profiles of read laser subpulses (see
Eq.(\ref{12})). It should be noted that four level scheme is more
general in the sense that it allows to adjust the anti-Stokes
photon frequency, at the same time this does not complicate the
calculations.

\begin{figure}[htb!]
 \centering
 \includegraphics[scale = 0.8]{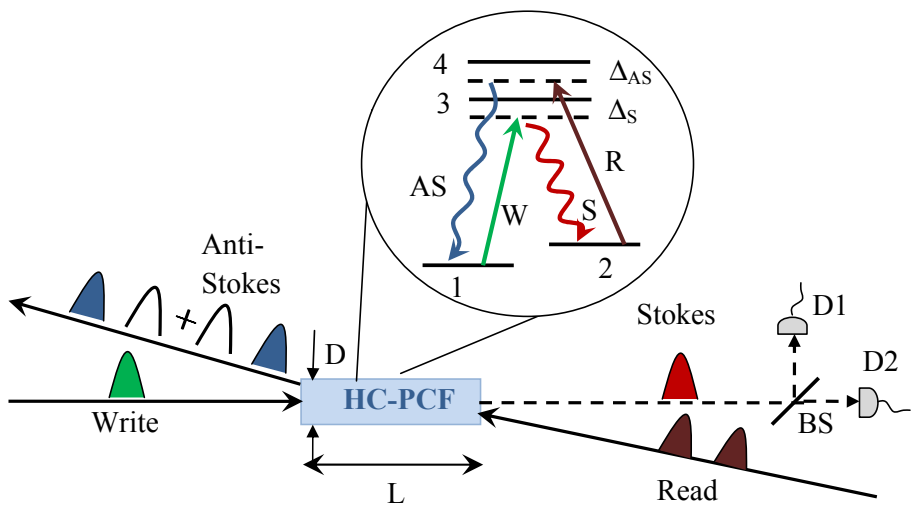}
 \caption{(Color online) Generation of Stokes and anti-Stokes
photons by a write W (green) and read R (brown) pulses. Setup and
atomic level scheme are in the inset). The case of two read pulses
and, correspondingly, the delocalization of anti-Stokes photon in
two possible modes (blue filled) is shown. An ensemble of cold
atoms is trapped inside a HC-PCF of length $L$ and diameter $D$.
The Stokes photon is detected with photon counters $D1$ and $D2$.}
\end{figure}

We consider a low-finesse cavity with damping rate $\chi=c/L$,
where Stokes and anti-Stokes cavity fields at frequencies $\omega
_{1}$ and\ $\omega _{2}$ are expressed, respectively, in terms of
annihilation (creation) operators $\hat a_{S}(\hat
a_{S}^\dagger)$\ \ and\ $\hat a_{AS}(\hat a_{AS}^\dagger)$ as
follows:
\begin{equation}
E_{i}^{(+)}(z,t)=\biggl ( \frac{2\pi \hbar \omega _{i}}{V}\biggr
)^{1/2}\hat a_{i}\exp(ik_{i}z-i\omega _{i}t),\  i=S,AS, \label{1}
\end{equation}%
with the quantization volume $V$ equal to the interaction volume
$V=\pi D^2 L/4$. Owing to the large detunings $\Delta_{W,R}$ we
adiabatically eliminate the upper states $\mid 3\rangle $ and
$\mid 4\rangle $, which are different generally, and then, in the
rotating frame approximation, the interaction Hamiltonian for the
total system in terms of the atomic collective spin operators
\begin{equation}
S^{\dag}=\frac{1}{\sqrt{N}}\sum\limits_{i=1}^{N}\sigma_{21}^{(i)},
\ \ S=(S^{\dag})^{\dag} \label{2}
\end{equation}
is written as \cite{19}
\begin{equation}
H=\hbar \sqrt{N}\bigl[ G(t)S^\dagger \hat
a_{S}^{\dagger}-F(t)S^\dagger \hat a_{AS} \bigr]+h.c.  \label{3}
\end{equation}%
In Eq.(\ref{2}) the summation is taken over all atoms, $\sigma
_{\alpha \beta}^{(i)}=\mid \alpha \rangle_{i}\langle \beta \mid $
is the atomic spin-flip operator in the basis of the two ground
states $\mid 1\rangle $ and $\mid 2\rangle $ for the $i$-th atom
and
\begin{eqnarray}
G(t)=g_{S}\frac{\Omega _{W}}{\Delta _{W}}f_{W}(t),   \label{4} \\
F(t)=g_{AS}\frac{\Omega _{R}(t)}{\Delta _{R}},  \label{5}
\end{eqnarray}
where $f_{W}(t)$ is the profile of the write pulse, while
$\Omega_R(t)$ is defined as
\begin{equation}
\Omega _{R}(t)=\sum\limits_{i=1}^{J}\Omega_if_i(t-t_i),  \label{6}
\end{equation}
showing that the $J$ well-separated readout pulses with peak Rabi
frequencies $\Omega_i$ are localized at time moments  $t_J
>t_{J-1}... >t_1$  with temporal profiles $f_i(t-t_i)$ and
relative delay between them much larger than their lengths  $T_i$.
Hereafter, for simplicity, we consider the Rabi frequency of the
write pulse real. In Eqs.(\ref{3}),(\ref{4}) the atom-photon
coupling constants are given by
\begin{equation}
\ g_{S}={2\pi \omega_{S} \overwithdelims() V}^{1/2}\mu _{32},
\quad  g_{AS}= {2\pi \omega_{AS} \overwithdelims() V}^{1/2}\mu
_{41},  \label{7}
\end{equation}%
with $\mu _{ij}$ the $\mid i\rangle \rightarrow \mid j\rangle $
transition dipole matrix element. Note that the Stark shifts of
the ground levels $\mid 1\rangle $ and $\mid 2\rangle $ induced by
the write and read pulses are considered smaller as compared to
the spectral width of the Stokes and anti-Stokes fields and can be
incorporated into their frequencies.

Our aim is to calculate the fluxes of Stokes and anti-Stokes
photons from the medium
\begin{equation}
\frac{dn_{i}}{dt} = \langle\hat a_{i,\mbox{out}}^{\dag}(t)\hat
a_{i,\mbox{out}}(t)\rangle, \quad i=S,AS, \label{8}
\end{equation}
where the annihilation operator of $i$-th output photon $\hat
a_{i,\mbox{out}}(t)$ is connected with the intracavity $\hat
a_{i}(t)$ and input $\hat a_{i,\mbox{in}}(t)$ annihilation
operators by the input-output relation $\hat
a_{i,\mbox{out}}(t)=\hat a_{i,\mbox{in}}(t)+\sqrt{2\chi}\hat
a_{i}(t)$ and satisfies the commutation relation
$[a_{i,\mbox{out}}(t),a_{i,\mbox{out}}^{\dag}(t^{\prime
})]=[a_{i,\mbox{in}}(t),a_{i,\mbox{in}}^{\dag}(t^{\prime})]=\delta
(t-t^{\prime })$. This technique is developed in our earlier work
\cite{19}. Here we sketch the main steps leading to the final
result.

The elimination of quantum fields in the bad cavity limit $\chi\gg
g_{S,AS}$ leads to the following links between the field and
atomic operators
\begin{eqnarray}
\hat a_{S}(t) =-i\sqrt{N}\frac{G(t)}{\chi
}S^{\dag}(t)-\sqrt{\frac{2}{\chi }}\hat a_{S,\mbox{in}}(t),
\label{9}\\ \hat a_{AS}(t) =-i\sqrt{N}\frac{F(t)}{\chi }S(t)\
-\sqrt{\frac{2}{\chi }}\hat a_{AS,\mbox{in}}(t), \label{10}
\end{eqnarray}
substitution of which into Eq.(\ref{8}) with input-output relation
yields for the photon fluxes
\begin{eqnarray}
\frac{dn_{S}}{dt} =\alpha (t)[N_{\mbox{sp}}(t)+1],  \label{11} \\
\frac{dn_{AS}}{dt} =\beta (t)N_{\mbox{sp}}(t),  \label{12}
\end{eqnarray}
where
\begin{eqnarray}
\alpha (t) =\frac{2N}{\chi}G^{2}(t),  \label{13} \\ \beta
(t)=\frac{2N}{\chi}\mid F(t)\mid^{2}  \label{14}
\end{eqnarray}
are the gains of Stokes and anti-Stokes fields, respectively, and
$N_{\mbox{sp}}=\left\langle S^{\dag}S\right\rangle$ is the number
of atomic spin-wave excitations having the form \cite{19}
\begin{equation}
N_{\mbox{sp}}(t)=\int_{-\infty}^t dt^{\prime }\alpha (t^{\prime
})\exp {\int_{t^\prime }^t d\tau [\alpha (\tau )-\beta (\tau
)-\Gamma_{\mbox{tot}}(\tau )]} \label{15}
\end{equation}
Here the total relaxation $\Gamma _{\mbox{tot}}(\tau
)=\gamma_c+\Gamma_W(t)+\Gamma_R(t)$ comprises the decay rate
$\gamma_c$ of atomic ground states coherence and the rates of
optical pumping between the states 1 and 2 via write and read
pulses
\begin{equation}
\Gamma _{W}(t)=\frac{\Omega _{W}^{2}}{\Delta
_{W}^{2}}f_{W}^2(t)\gamma _{32}, \quad \Gamma
_{R}(t)=\frac{\mid\Omega _{R}(t)\mid^{2}}{\Delta
_{R}^{2}}\gamma_{41} \label{16}
\end{equation}
\begin{figure}[b] \centering
{\includegraphics* [scale =0.7]{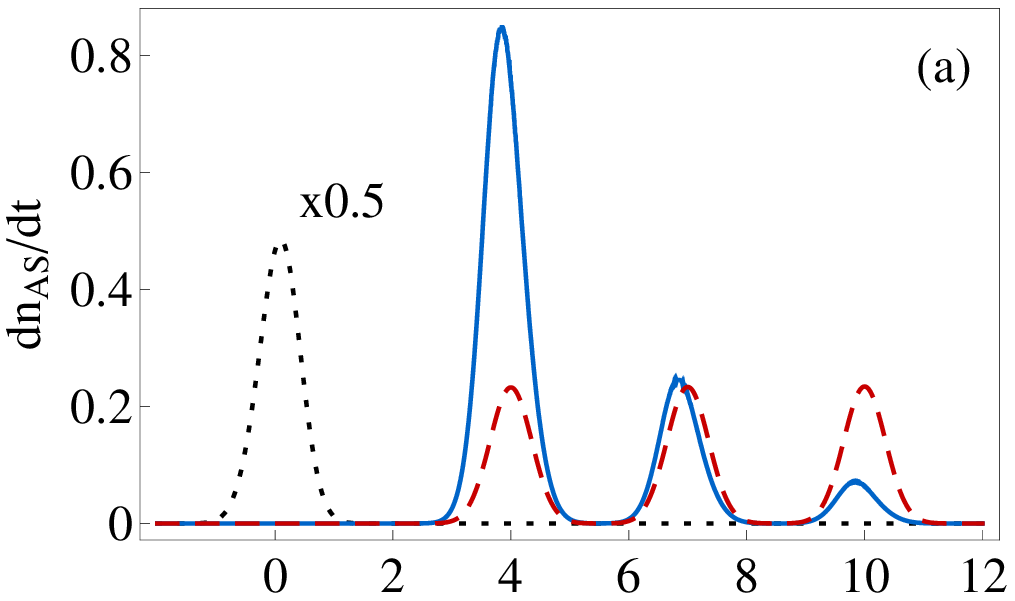}}\\
{\includegraphics* [scale =0.7]{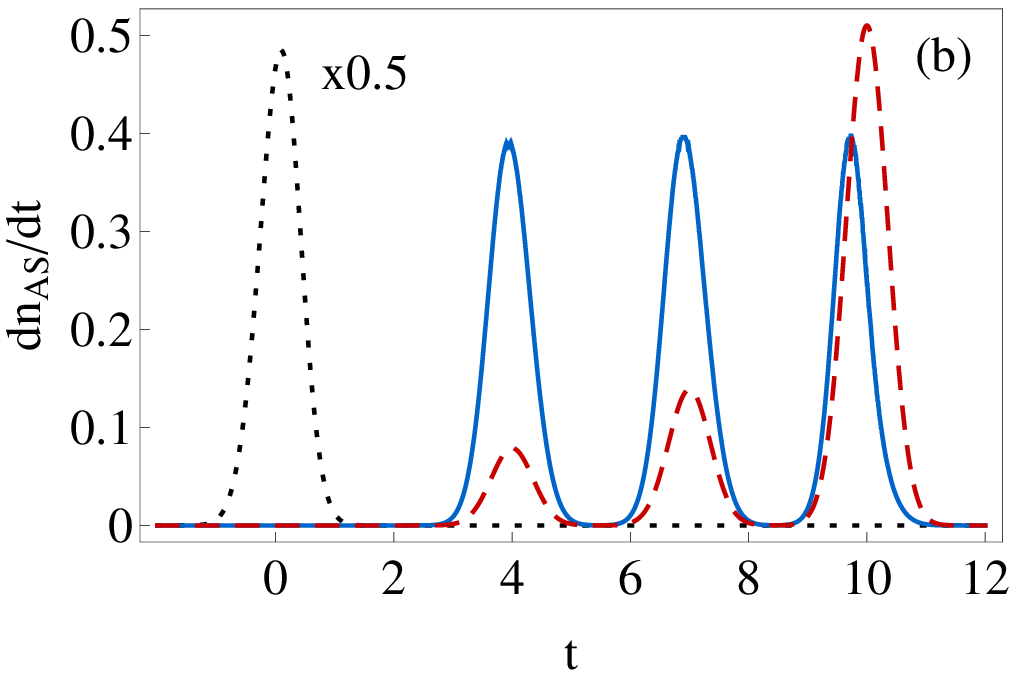}} \caption{(Color
online) Anti-Stokes photon flux as a function of time in units of
$\gamma$ in two cases of: (a) three identical read pulses (red
dashed lines) and (b) equal intensities of three temporal modes of
anti-Stokes field (blue solid lines). Black dotted line shows the
Stokes pulse of unit area. The sum area of anti-Stokes pulses is
0.98. In both cases the Stokes and read pulses are scaled by
factors 0.5 and 0.08, respectively. \label{Figs2}}
\end{figure}
where $\gamma _{ij}$ is a partial decay rate of upper level $i$ to
the state $j$ giving in sum the spontaneous decay rate of upper
states $\gamma=\gamma_i=\sum _j\gamma _{ij}$ .

There are two important consequences coming out from
Eqs.(\ref{11}),(\ref{12}) and (\ref{15}). First, for the large
signal-to-noise ratio $\alpha/\Gamma_W\gg 1$ and
$\beta/\Gamma_R\gg1$, the relaxation terms in Eq.(\ref{15}) can be
neglected. Then, the photon numbers are readily found from
Eqs.(\ref{11}),(\ref{12}) and (\ref{15}) to be
\begin{eqnarray}
n_{S}(t)=N_{\mbox{sp}}(t)=e^{\int_{-\infty}^t d\tau \alpha (\tau
)}-1, \quad t\leq T_W   \label{17} \\
n_{AS}(t)=n_{S}(\infty)(1-e^{-\int_{-\infty}^t d\tau \beta (\tau
)}), \quad t\geq T_W+\tau_D \label{18}
\end{eqnarray}
showing that, on the one hand, the number of detected Stokes
photons equals to the number of spin excitations stored in the
atomic medium during the write pulse of duration $T_W$ and, on the
other hand, it unambiguously determines the number of anti-Stokes
photons retrieved from the medium at the end of the read
subpulses. Here we have used that the time delay $\tau_D$ between
the read and write laser pulses is large compared to $T_W$ and, at
the same time, it is much shorter than the spin decoherence time
$\gamma_c^{-1}$.

Second, taking into account the orthogonality of the functions
$f_i(t-t_i)$ in Eq.(\ref{6}), the Eq.(\ref{12}) describes the
deterministic generation of the anti-Stokes photon entangled in
nonoverlapping temporal modes with controllable intensities. An
important issue arising here is the phase-coherence of anti-Stokes
photon delocalization in multiple time bins. In the next section
we suggest an experimental test to verify this coherence.

For numerical calculations we consider cold $^{87}$Rb atoms with
the ground states $5S_{1/2}(F=1,2)$ and excited states
$5P_{1/2}(F=2)$ and $5P_{3/2}(F=2)$ as the atomic states 1, 2, and
3,4 in Figure 1, respectively. Number of atoms confined in a
hollow-core fiber of the length $L\sim 3$cm and diameter $D\sim
5\mu$m is about $N\sim10^4$ \cite{22}, the fields are tuned far
from the one-photon resonance by $\Delta_{W,R}=20\gamma$. Below
$\gamma _c$ is neglected. The durations of the write and read
subpulses in Eq.(\ref{6}) are taken $T_W\sim T_i\sim 1\mu$s. Here
we assume that anti-Stokes photons are retrieved and detected with
efficiency close to unity taking into account the recent progress
in this area \cite{23,24}. In Figure 2 we present the retrieved
anti-stokes photon for two sets of read pulses. In the first case,
the amplitudes of three read pulses are the same [Figure 2(a)],
while in the second one they are redesigned such that the temporal
modes of the anti-stokes photon have equal intensities [Figure
2(b)]. The number of anti-stokes photons in the modes is
determined by the areas of the corresponding peaks. The parameters
chosen for the write pulse are such that in average only one
Stokes photon is generated: $n_{S}(\infty)\sim 1$ or, equivalently
from Eq.(\ref{17}), $\int_{-\infty}^{\infty} d\tau \alpha (\tau
)\sim 0.7$.

\section{Phase coherence of atomic spin-wave conversion into multi-pulse anti-Stokes photon}

The pure state of the anti-Stokes photon entangled over $J$
temporal modes has the form
\begin{equation}
\mid \Psi_{AS}\rangle=\sum_{j=1}^J C_{j}\mid
1\rangle_{j}\prod_{i\neq j}^{J}\mid0\rangle_{i}, \label{19}
\end{equation}
where $|1\rangle_{j}$ and $|0\rangle_{j}$ denote Fock states with
zero and one photon, respectively, at the time $t_j$. Here the
complex amplitudes $C_ i$ with normalization $\sum_j |C_j|^2 = 1$
depend on the intensities and, in the ideal case, only on relative
phases of readout pulses. The main requirement is that these
pulses should maintain mutual coherence to preserve entanglement
between temporal modes because the entanglement implieas a
constant phase relation between its different components. However,
the collective conversion of the atomic spin excitation into
anti-Stokes photons can introduce uncontrollable phases
$\varphi_{\mbox{random}}$ due to, for example, the thermal motion
of the atoms, resulting in decoherence in the state (19). Here we
suggest an experimental test to verify that the coherence is
preserved during the conversion process.

\begin{figure}\centering
{\includegraphics* [scale =0.55]{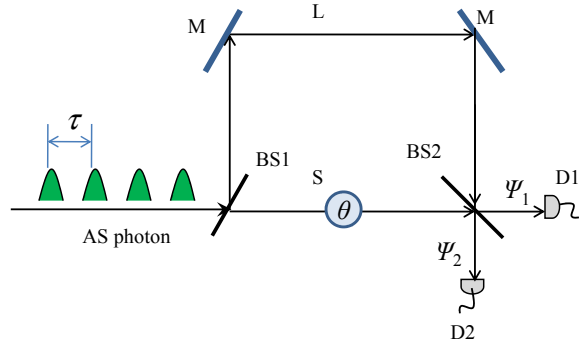}} \caption{(Color
online) Schematic setup of coherence detection between consecutive
temporal modes of anti-Stokes field. Passing the long ($L$) and
short ($S$) arms of the interferometer the anti-Stokes photon
interferes with itself in observing signals on the detectors $D1$
and $D2$. $\tau$ is the time separation between the modes, $\mid
\Psi_{1,2}\rangle$ are the output states of the ani-Stokes photon,
M- mirrors, BS - beam  splitters, $\theta$ - phase shifter.
\label{Fig3}}
\end{figure}
The single anti-Stokes photon delocalized in multi time-bins is
sent through a Franson-type interferometer with a phase shifter
$\theta$ inserted in one of the arms (Figure 3). For simplicity,
we consider the case, when all readout pulses in Eq.(\ref{6}) are
equally separated in time. Then the length difference of long and
short arms is supposed to match the time separation $\tau$ between
two consecutive readout pulses $L_{\mbox{long}}-L_{\mbox{short}} =
c \tau$, so that the outputs of the beam-splitter $BS2$ could
interfere by observing signals on the detectors $D1$ and $D2$
showing fringe pattern in dependence on $\theta$. As the
visibility of the pattern is sensitive to random phases acquired
by anti-Stokes temporal modes, this allows one to test the
coherence of multi-pulse conversion of the atomic spin excitation.
To show this, we calculate the number of anti-Stokes photons
measured by $D1$ and $D2$  detectors within a time interval much
longer than the photon total duration
\begin{equation}
n_{AS}^{(i)} = \int_{-\infty}^{\infty} dt \langle\Psi_i\mid \hat
a^{\dag}(t)\hat a(t)\mid \Psi_i\rangle, \quad i=1,2, \label{20}
\end{equation}
where $\mid \Psi_i\rangle, i=1,2,$ is the ani-Stokes photon state
at $i$-th output port of the beam-splitter $BS2$. Omitting the
vacuum part, which does not contribute to Eq. (\ref{20}), these
states are given by
\begin{equation}
\mid\Psi_{1,2}\rangle=\frac{1}{2}(e^{i\theta}\mid\Psi_{AS}^{(s)}\rangle
\pm \mid\Psi_{AS}^{(l)}\rangle), \label{21}
\end{equation}
where $\mid\Psi_{AS}^{(s,l)}\rangle$ are defined in Eq. (\ref{19})
with $j$-th mode single-photon states
\begin{equation}
|1^{(s)}\rangle_j = \int_{-\infty}^{\infty}\Phi_j(t-t_j)\hat
a_j^\dag (t)\mid 0\rangle dt,\label{22a}
\end{equation}
\begin{equation}
|1^{(l)}\rangle_j = \int_{-\infty}^{\infty}\Phi_j(t-t_j-\tau)\hat
a_j^\dag (t)\mid 0\rangle dt,\label{22}
\end{equation}
respectively. Here the real functions $\Phi_j(t-t_j)$ form an
orthonormal set of temporal modes localized around $t=t_j:$
\begin{equation}
\int_{-\infty}^{\infty}\Phi_j(t-t_j)\Phi_k(t-t_k)dt=\delta_{jk}.
\label{23}
\end{equation}
Substituting Eqs.(\ref{21}) and (\ref{22a}), (\ref{22}) into
Eq.(\ref{20}) and using $\hat a(t)=\sum_{j=1}^J \hat a_j(t)$ one
obtains
\begin{eqnarray}
n_{AS}^{(1,2)} &=& \frac{1}{2}\biggr [1\pm \sum_{j,k}^J
\cos(\theta +\Delta\varphi_{j,k})\mid C_j\mid \mid C_k\mid
\nonumber \\ &\times
&\int_{-\infty}^{\infty}\Phi_j(t-t_j)\Phi_k(t-t_k-\tau)dt \biggr
], \label{24}
\end{eqnarray}
where $\Delta\varphi_{j,k}$ is the relative phase between the $j$
and $k$ temporal modes. From this equation we recognize that the
fringe pattern visibility depends on photon numbers
$n_{i,k}=|C_{i,k}|^2$ in $j$ and $k$ modes and on the integral
overlap of these modes which, according to Eq.(\ref{23}), is
finite only if $j=k+1$. For our purposes, the simplest case of
equal photon numbers $n_j=1/J$ in all modes and identical mode
functions $\Phi_j(t-t_j)\equiv \Phi(t-t_j), j=1,...,J$, shown as
an example in Figure 2b, suffices to illustrate the phase
coherence across all time bins. With these simplifications, we
have
\begin{equation}
n_{AS}^{(1,2)} = \frac{1}{2}[1\pm \frac{1}{J}\sum_{k}^{J-1}
\cos(\theta+\Delta\varphi_{k+1,k})]. \label{25}
\end{equation}
As the unknown phases that would originate from the atomic
spin-wave conversion are time-independent, the photon numbers
averaged over $\varphi_{\mbox{random}}$ distribution, for real
Rabi frequencies $\Omega_i$ in Eq.(\ref{6}), are finally obtained
as
\begin{equation}
\overline{n_{AS}^{(1,2)}} = \frac{1}{2}[1\pm
\frac{J-1}{J}\overline{\cos(\theta+\varphi_{\mbox{random}})}].
\label{26}
\end{equation}
\begin{figure}[t] \centering {\includegraphics*
[scale =0.7]{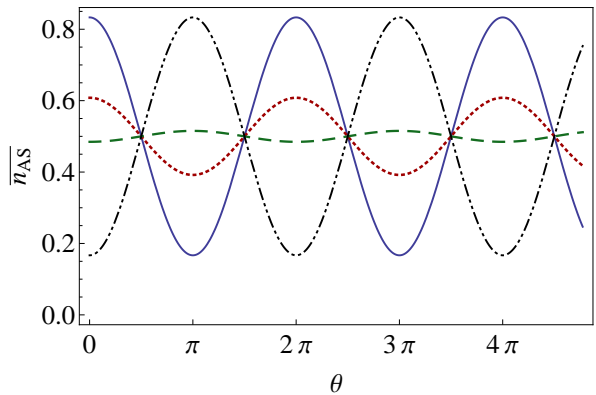}} \caption{(Color online) Averaged
anti-Stokes photon number $\overline{n_{AS}^{(1)}}$ as a function
of $\theta$ for three different variances of
$\varphi_{\mbox{random}}$: 0 (blue solid line),  1.5 (red dotted
line), 3 (green dashed line). Black dash-dotted line shows
$\overline{n_{AS}^{(2)}}$ for zero variance. \label{Fig4}}
\end{figure}
In Figure 4 we present $\overline{n_{AS}^{(1,2)}}$ found
numerically for Gaussian distribution of $\varphi_{\mbox{random}}$
with zero mean value and different variances demonstrating the
dependence of interference contrast on phase scattering, the more
this variation, the smaller the visibility. Therefore, the
proposed method to observe the interference between temporal modes
with maximal visibility $(J-1)/J$ can be used to test the
coherence of multi-time-bins entanglement of a single anti-Stokes
photon, which is deterministically generated from a stored single
atomic spin-excitation.

\section{Conclusions}

In conclusion, we have proposed a highly efficient heralded source
of anti-Stokes single-photons entangled in multiple temporal
modes. The source is based on the heralded creation of one atomic
spin excitation followed by deterministic conversion of the latter
into single anti-Stokes photon that is delocalized in
multi-time-bins. With experimentally verified heralded creation of
a single atomic spin excitation, the source clearly provides high
purity of single-photon states. The waveforms of anti-Stokes
temporal modes are controlled by the shape of read laser pulses,
while the phase coherence across all time bins can be
experimentally verified by the suggested mechanism. Such
controlled scheme can be used first of all for implementation of
quantum repeaters based on time-bin entangled single-photon
states.

\bigskip
\subsection*{Acknowledgments}

This research has been conducted in the scope of European Union
Seventh Framework Programme Grants No. GA-295025-IPERA and
No.609534-SECURE-R2I, and ANSEF Grant PS-opt 3201. We acknowledge
additional support from the International Associated Laboratory
(CNRS-France and SCS Armenia) IRMAS.

\end{document}